# Ferromagnetism in the Hubbard model on the square lattice: Improved instability criterion for the Nagaoka state *

P. Wurth, G. Uhrig, E. Müller-Hartmann

Institut für Theoretische Physik

Universität zu Köln, Zülpicher Straße 77, 50937 Köln

December 8, 1995

## Abstract

The instability of the fully polarized ferromagnetic state (Nagaoka state) with respect to single spin flips is re-examined for the Hubbard model on the square lattice with a large family of variational wave functions which include correlation effects of the majority spins in the vicinity of the flipped spin. We find a critical hole density of $\delta_{cr} = 0.251$ for $U = \infty$ and a critical coupling of $U_{cr} = 77.7t$. Both values improve previous variational results considerably.

---

*Research performed within the program of the Sonderforschungsbereich 341 supported by the Deutsche Forschungsgemeinschaft

# 1 Introduction

Since the introduction of the Hubbard model thirty years ago [1–3] an important question has been whether this simple model of correlated electrons on a lattice can provide some insight in the mechanism for ferromagnetism in transition metals. A first proof of the stability of the fully polarized ferromagnetic state (Nagaoka state) at $T = 0$ was presented by Nagaoka [4] for exactly one additional electron added to the half filled band: For infinite intraatomic Coulomb repulsion $U$ the Nagaoka state was found to be the unique ground state on most common lattices. Unfortunately, an extension of Nagaoka's result to thermodynamically relevant finite densities of added particles (and holes for bipartite lattices) was never achieved. Various numerical [5, 6] and analytical [7] calculations showed that the ground state in the case of two holes on the square lattice is a state with lower total spin. Barbieri et al. did demonstrate local stability of the Nagaoka state for larger numbers of holes, but their result was still confined to vanishing hole density in the thermodynamic limit [8]. A proof of the stability of the Nagaoka state for finite hole densities has only recently been obtained for special lattices [9, 10] or a special choice of the hopping matrix $t_{\mathbf{x},\mathbf{y}}$ [11]. On infinite dimensional lattices, the region of local stability of the Nagaoka state was recently calculated exactly from the lower band edge of the Green function of the flipped electron [12].

In view of the difficulty of proving the stability of the fully polarized ferromagnetic state for common lattices, in recent years a number of variational wave functions have been investigated in order to establish a region in the phase diagram, where the Nagaoka state is definitely unstable. This region can be charcterized by a critical hole density $\delta_{cr}$ above which the Nagaoka state is unstable at $U = \infty$ and by a critical Coulomb repulsion $U_{cr}$ below which the Nagaoka state is unstable for all hole densities. The present paper is devoted to an extension of the variational analysis of the instability region in the case of the square lattice which is by far the most extensively studied lattice in the present context.

The first variational result of $\delta_{cr} = 0.49$ was obtained by Shastry, Krishnamurthy and Anderson (SKA) [13] who used a simple Gutzwiller projected single spin flip variational wave function. This result was improved to $\delta_{cr} = 0.41$ by Basile and Elser [14] who investigated on finite lattices a wave function proposed earlier by Roth [15]. The same wave function was recently analyzed exactly in the thermodynamic limit on various lattices [16].

Two different approaches resulted in a further reduction of the critical hole density to the value of $\delta_{cr} = 0.29$. First von der Linden and Edwards [17] used a variational wave function on finite lattices that consists of a coherent superposition of states, where the flipped spin is fixed at one site and the wave function of the majority spins is an optimal Slater determinant of one particle orbitals. Without the movement of the flipped spin this corresponds to the ansatz used earlier by Richmond and Rickayzen [18]. Von der Linden and Edwards obtained $\delta_{cr} = 0.29$ and $U_{cr} = 42t$. Later Hanisch and Müller-Hartmann [19] studied an ansatz which took into account various local correlations in addition to the ansatz of SKA [13] and of Gebhard and Zotos [20]. This type of wave function is more local than the ansatz of von der Linden and Edwards – and therefore can be evaluated in the thermodynamic limit – but is able to include correlation effects ignored by von der Linden and Edwards. Hanisch and Müller-Hartmann obtained $\delta_{cr} = 0.29$ and $U_{cr} = 63t$.

Exact diagonalization of finite clusters points at a lower critical hole density of $\delta_{cr} = 0.195$ [21], similar to a result of density matrix renormalization group calculations by Liang and Pang who obtained $\delta_{cr} = 0.22$ [22]. Putikka et al. found a ground state with reduced total spin for all hole densities by extrapolating a high temperature expansion of the Helmholtz free energy [23]. Because of the uncertainties of their extrapolation to $T = 0$, especially for low hole densities, their calculation cannot be considered a definite proof of the complete absence of a Nagaoka ground state.

In this paper we present extensions of both the ansatz used in [19] and a spin wave ansatz that we studied previously [24], in order to reduce the remaining gap between the results of numerical calculations on finite clusters and the best variational bounds. In the second section we will give a short outline of our method. The following section contains the results of our investigation which are finally discussed in the last section.

## 2 Variational wave functions and calculation of the energy

We study the Hubbard model

$$\mathcal{H} = - \sum_{<\mathbf{xy}>} t\left(c^+_{\mathbf{y}\uparrow}c_{\mathbf{x}\uparrow} + c^+_{\mathbf{y}\downarrow}c_{\mathbf{x}\downarrow}\right) + U \sum_{\mathbf{x}} n_{\mathbf{x}\uparrow}n_{\mathbf{x}\downarrow} \qquad (1)$$

on the square lattice at $T = 0$ in the thermodynamic limit. The sum $\sum_{<\mathbf{xy}>}$ is confined to nearest neighbor sites $\mathbf{x}$ and $\mathbf{y}$. We examine two classes of variational wave functions in order to investigate the stability of the fully polarized ferromagnetic state (Nagaoka state)

$$|\mathcal{N}\rangle = \prod_{\{\mathbf{k}|\epsilon_{\mathbf{k}} \leq \epsilon_F\}} c^+_{\mathbf{k}\uparrow}|0\rangle \qquad (2)$$

with respect to single spin flips.

The first class of wave functions was studied previously by Hanisch and Müller-Hartmann [19] and has the general form

$$|\mathbf{q}\rangle_{\mathbf{k}_F} = \frac{1}{\sqrt{L}} \sum_\alpha \sum_\mathbf{m} e^{i\mathbf{qm}} \sum_\alpha \psi_\alpha c^+_{\mathbf{m}\downarrow} \mathcal{A}_{\mathbf{m},\alpha} c_{\mathbf{k}_F\uparrow} |\mathcal{N}\rangle \qquad (3)$$

with variational parameters $\psi_\alpha$. A spin-$\uparrow$ electron is removed from the Fermi surface and the spin-$\downarrow$ electron is created at the band bottom ($\mathbf{q} = \mathbf{0}$) to achieve the largest possible gain in kinetic spin-$\downarrow$ energy. The operators $\mathcal{A}_{\mathbf{m},\alpha}$ are products of local particle-hole excitations of the majority spin-$\uparrow$ electrons within a finite distance from the position $\mathbf{m}$ of the flipped spin and describe the dynamical deformation of the spin-$\uparrow$ Fermi sea around the flipped spin. Because of the translational invariance of the system we assume that the operators $\mathcal{A}_{\mathbf{m},\alpha}$ are obtained from $\mathcal{A}_{\mathbf{0},\alpha}$ by translation. The momentum carried by the excited state $|\mathbf{0}\rangle_{\mathbf{k}_F}$ is $-\mathbf{k}_F$ where $\mathbf{k}_F$ can be any Fermi surface momentum vector. The SKA-ansatz [13] as an example of the ansatz (3) corresponds to the choice of two local operators $\mathcal{A}_{\mathbf{0},1} = c_{\mathbf{0}\uparrow} c^+_{\mathbf{0}\uparrow}$ and $\mathcal{A}_{\mathbf{0},2} = 1$. Variational wave functions of this type have been used in [19], and more recently in [25] to determine the phase diagram of various two-dimensional lattices. The energy of an ansatz containing all one particle-hole pair operators of the form $\mathcal{A}_{\mathbf{0},(1,0)} = c_{\mathbf{1}\uparrow} c^+_{\mathbf{0}\uparrow}$ as well as $\mathcal{A}_{\mathbf{0},(0,1)} = c_{\mathbf{0}\uparrow} c^+_{\mathbf{1}\uparrow}$ was recently calculated by Uhrig and Hanisch [16] on various lattices. In the present paper we use an efficient numerical algorithm which enables us to increase the number of $\mathcal{A}_{\mathbf{m},\alpha}$ operators which was 29 in [19] to more than 1000 including operators $\mathcal{A}_{\mathbf{m},\alpha}$ with up to 3 particle-hole excitations.

The second class of variational wave functions we consider allows for bound states between the flipped spin and the hole in the majority spin Fermi sea. This additional freedom leads to an infinite number of variational parameters $\phi_{\mathbf{k},\alpha}$ with $\mathbf{k} \epsilon BZ$:

$$|\mathbf{q}\rangle = \frac{1}{\sqrt{L}} \sum_\mathbf{k} \sum_\mathbf{m} e^{i(\mathbf{k+q})\mathbf{m}} \sum_\alpha \phi_{\mathbf{k},\alpha} c^+_{\mathbf{m}\downarrow} \mathcal{A}_{\mathbf{m},\alpha} c_{\mathbf{k}\uparrow} |\mathcal{N}\rangle. \qquad (4)$$

In [24] we showed, that in the vicinity of $\mathbf{q} = (0, \pi)$ the binding of a hole leads to a significant reduction of the critical hole density compared to the corresponding scattering states containing the same local correlation terms $\mathcal{A}_{\mathbf{m},\alpha}$. The latter calculation included variational parameters $\phi_{\mathbf{k},\alpha}$ for infinitely many values of $\mathbf{k}$ and for 5 values of $\alpha$. It turned out that an extension of this calculation to more $\mathcal{A}$-operators is numerically tedious. However, since the best wave function is a bound state of spin-$\uparrow$ hole and spin-$\downarrow$ electron the calculation can be simplified by only considering holes that are localized in the vicinity of the flipped spin. The corresponding ansatz uses site representation and has the form

$$|\mathbf{q}\rangle = \frac{1}{\sqrt{L}} \sum_{\mathbf{m}} \sum_{\mathbf{n}} e^{i\mathbf{q}\mathbf{m}} \sum_{\alpha} \tilde{\psi}_{\mathbf{n},\alpha} c^+_{\mathbf{m}\downarrow} \mathcal{A}_{\mathbf{m},\alpha} c_{\mathbf{m}+\mathbf{n}\uparrow} |\mathcal{N}\rangle \qquad (5)$$

$$= \frac{1}{\sqrt{L}} \sum_{\mathbf{m}} e^{i\mathbf{q}\mathbf{m}} \sum_{\alpha} \psi_{\alpha} c^+_{\mathbf{m}\downarrow} \tilde{\mathcal{A}}_{\mathbf{m},\alpha} |\mathcal{N}\rangle, \qquad (6)$$

where $\mathbf{n}$ are lattice sites with a finite distance from the origin and $\tilde{\mathcal{A}}_{\mathbf{m},(\alpha,\mathbf{n})} = \mathcal{A}_{\mathbf{m},\alpha} c_{\mathbf{m}+\mathbf{n}\uparrow}$. The energy of the spin wave states (6) can be obtained by the same algorithm that is used to calculate the energy of the corresponding scattering states (3).

Variation of the difference between the energy of the Nagaoka state and the energies of the states (3) and (6), respectively, leads to the generalized eigenvalue problem

$$\sum_{\beta} \mathcal{L}_{\alpha,\beta} \psi_{\beta} = \epsilon \sum_{\beta} \mathcal{P}_{\alpha,\beta} \psi_{\beta}. \qquad (7)$$

The spin flip energy $\epsilon$ is given by the lowest eigenvalue and is calculated numerically. The matrices involved are

$$\mathcal{P}_{\alpha,\beta} = \langle \mathcal{N} | \mathcal{A}^+_{\mathbf{0},\alpha} \mathcal{A}_{\mathbf{0},\beta} | \mathcal{N} \rangle \qquad (8)$$

and

$$\mathcal{L}_{\alpha,\beta} = \sum_{\mathbf{m}} e^{-i\mathbf{q}\mathbf{m}} \langle \mathcal{N} | \mathcal{A}^+_{\mathbf{m},\alpha} c_{\mathbf{m}\downarrow} [\mathcal{H}, c^+_{\mathbf{0}\downarrow} \mathcal{A}_{\mathbf{0},\beta}] | \mathcal{N} \rangle - \epsilon_F \mathcal{P}_{\alpha,\beta} \qquad (9)$$

for the scattering states (3). In the case of the spin waves (6) they can be written as

$$\mathcal{P}_{\alpha,\beta} = \langle \mathcal{N} | \tilde{\mathcal{A}}^+_{\mathbf{0},\alpha} \tilde{\mathcal{A}}_{\mathbf{0},\beta} | \mathcal{N} \rangle. \qquad (10)$$

and

$$\mathcal{L}_{\alpha,\beta} = \sum_{\mathbf{m}} e^{-i\mathbf{q}\mathbf{m}} \langle \mathcal{N} | \tilde{\mathcal{A}}^+_{\mathbf{m},\alpha} c_{\mathbf{m}\downarrow} [\mathcal{H}, c^+_{\mathbf{0}\downarrow} \tilde{\mathcal{A}}_{\mathbf{0},\beta}] | \mathcal{N} \rangle \qquad (11)$$

The commutators of the local operators and the Hamiltonian and the resulting expectation values are computed and transformed into normal order algebraically using a C-program. Since the Nagaoka state contains no spin-$\downarrow$ particle the spin-$\downarrow$ electron operators only select a single term in the $\mathbf{m}$-summation of the matrices $\mathcal{L}$. The remaining expectation values of products of local spin-$\uparrow$ electron operators can be evaluated numerically using the identity

$$\langle \mathcal{N} | c_{\mathbf{x}_n} \cdots c_{\mathbf{x}_1} c^+_{\mathbf{y}_1} \cdots c^+_{\mathbf{y}_n} | \mathcal{N} \rangle = \det(\langle \mathcal{N} | c_{\mathbf{x}_i} c^+_{\mathbf{y}_j} | \mathcal{N} \rangle)_{\substack{i=1,n \\ j=1,n}}. \qquad (12)$$

The elements of the one particle density matrix $\langle \mathcal{N}|c_{\mathbf{x}_i}c^+_{\mathbf{y}_j}|\mathcal{N}\rangle$ are calculated once for a given hole density by using the recursion formulae given in [19]. Since the determinants (12) we have to evaluate have dimensions up to $n = 7$ it is most efficient to evaluate them by a (standard) Gaussian algorithm.

In [19] the ansatz (3) was investigated with up to 28 variational parameters. Since this ansatz already contained the most relevant terms $\mathcal{A}_{\mathbf{m},\alpha}$ a noticeable improvement was only possible by systematically including a large number of additional correlation terms. As it is not at all obvious which of the many operators $\mathcal{A}_{\mathbf{m},\alpha}$ should be included for a better variational wave function, a test was used in order to select the most promising terms. We started with the most general wave function of [19], added single correlation terms one after the other and compared the resulting energies at a given hole density close to the critical hole density. Those operators which gave the lowest energies were then included in our final ansatz. Our best wave function contained operators with up to 3 particle-hole excitations and it included local processes within a $9 \times 9$-plaquette around the flipped spin. The same strategy was used in order to improve the spin wave results [24] with the ansatz (6). In this case the total number of terms $\tilde{\mathcal{A}}_{\mathbf{0},(\alpha,\mathbf{n})}$ is given by the number of correlation terms $\mathcal{A}_{\mathbf{0},\alpha}$ multiplied by the number of allowed positions $\mathbf{n}$ of the hole. The corresponding matrix elements (12) then contain one additional particle-hole pair. In addition to this increase of numerical efforts the symmetry of this ansatz is also reduced compared to the scattering states since the best spin wave is found at the zone boundary with $\mathbf{q} = (0, \pi)$. As a consequence we were able to include many more terms $\mathcal{A}_{\mathbf{0},\alpha}$ in the scattering state (3) than in the spin wave state (6).

## 3 Results

Our most general wave function (3) contained 1100 correlation terms $\mathcal{A}_{\mathbf{0},\alpha}$. As described above we chose the most promising terms from all $\mathcal{A}_{\mathbf{0},\alpha}$ containing up to two particle-hole excitations with spin-↑-operators in a $9 \times 9$-plaquette and three particle-hole excitations in a $3 \times 3$-plaquette centered at the flipped spin. Of the 1100 correlation terms 1, 22, 1045 and 32, contain zero, one, two and three particle-hole excitations, respectively. The resulting phase diagram is shown in figure 1 in comparison to the best result obtained in [19]. The global stability limit set by a phase separation [19] between a spin density wave at half filling and the Nagaoka state is also shown. The on site repulsion $U$ is represented in terms of $U_{red} = U/(U + U_{BR})$ with the Brinkman-Rice critical coupling $U_{BR} = \frac{128}{\pi^2}t$ [26] as a natural energy reference. We obtain a critical hole density of $\delta_{cr} = 0.251$ and a critical coupling of $U_{cr} = 77.7t$. This is a remarkable improvement in comparison to the best values given so far of $\delta_{cr} = 0.29$ [17, 19] and $U_{cr} = 63t$ [19] and the variational critical hole density is now quite close to the numerical estimates of $\delta_{cr} = 0.195$ [21] and $\delta_{cr} = 0.22$ [22].

We have also evaluated the ansatz (6) for various local correlation terms. For $\mathbf{q} = (0, \pi)$ all calculated spin wave energies lead to a reduction of the critical hole densities compared to scattering states containing the same correlations, in agreement with the results given in [24]. However the difference between the corresponding critical hole densities shrinks if one takes into account more and more correlations.

It turns out to be necessary to allow approximately 50 lattice sites $\mathbf{n}$ for the position of the spin-↑ hole in order to achieve a good approximation to the complete bound state ansatz (4). As a consequence we can include 50 times more local correlation terms in the scattering state than in the spin wave state with even less numerical effort. Thus it is numerically not feasible to calculate the energy of a spin wave corresponding to our best scattering state. Using a moderate number of correlation terms in our spin wave ansatz we are not able to improve the above results obtained with our most flexible scattering state, although the instability of the Nagaoka state is probably due to a bound spin wave state with $\mathbf{q} = (0, \pi)$.

## 4 Summary

We have re-investigated the stability of the fully polarized ferromagnetic ground state (Nagaoka state) with respect to single spin flips for the Hubbard model on the square lattice. Two classes of variational wave functions were used that contain local correlations of the majority spin-↑ electrons in the vicinity of the flipped spin: scattering states (3) and spin waves (6). In the scattering states a majority spin-↑ electron is removed from the Fermi surface and a spin-↓ electron is created at the band bottom. The spin wave states describe a bound state between the flipped spin and the hole in the Nagaoka state.

In both variational wave functions the correlations between the spin-↓ electron and the majority electrons were taken into account by including local operators $\mathcal{A}_{\mathbf{m},\alpha}$ describing particle-hole excitations in the vicinity of the flipped spin. We selected systematically those correlation terms which reduce the energy of the variational wave function most efficiently.

In principle, spin wave states with momentum $\mathbf{q} = (0, \pi)$ reduce the critical hole density further in comparison to scattering states containing the same correlation terms which confirms the scenario described in [24]. The numerical effort to calculate the variational energies is, however, much larger for the spin waves than for the scattering states. As a consequence we have not obtained the best estimate of the instability region with spin wave states, but rather with scattering states.

Our best scattering state containing 1100 correlation terms included operators $\mathcal{A}_{\mathbf{m},\alpha}$ with up to 2 spin-$\uparrow$ particle-hole excitations from a $9 \times 9$-plaquette centered at the flipped spin and correlation terms with 3 spin-$\uparrow$ particle-hole excitations confined to a $3 \times 3$-plaquette centered at the spin-$\downarrow$ electron. With this ansatz we obtained a critical hole density of $\delta_{cr} = 0.251$ and a critical coupling of $U_{cr} = 77.7t$ which improves the previous best estimates of $\delta_{cr} = 0.29$ and $U_{cr} = 63t$ [19]. The full regions of instability were shown in figure 1.

If the conclusion of Putikka et al. [23] that the ground state has a reduced total spin for all hole densities is correct some major correlations will still be missing in our variational states. The uncertainties of the low temperature extrapolation of their high temperature series would not be inconsistent with $\delta_{cr} \lesssim 0.1$ [27]. In contrast to this work the numerical estimates of Hirsch [21] and Liang and Pang [22], who obtain $\delta_{cr} = 0.195$ and $\delta_{cr} = 0.22$, respectively, from finite size extrapolations of exact spin flip states suggest that our local ansatz gives a quite satisfying description of the correlations between the flipped spin and the majority spins. Putikka et al. [23] however find a ground state with lower total spin for all hole densities. Their extrapolation of a high temperature expansion implies certain error bars at $T = 0$ especially for low hole densities and thus gives no definite prove of the instability of the Nagaoka state. Since our results only give upper bounds for the local stability of the Nagaoka state the true region of stability could however be even smaller.

In the work presented here, the progress in variationally estimating the instability region has been achieved by extensive use of computer algebra in terms of a fast C-program and by optimizing the numerical evaluations. Only minor further improvements will be possible by enhancing the computational efforts.

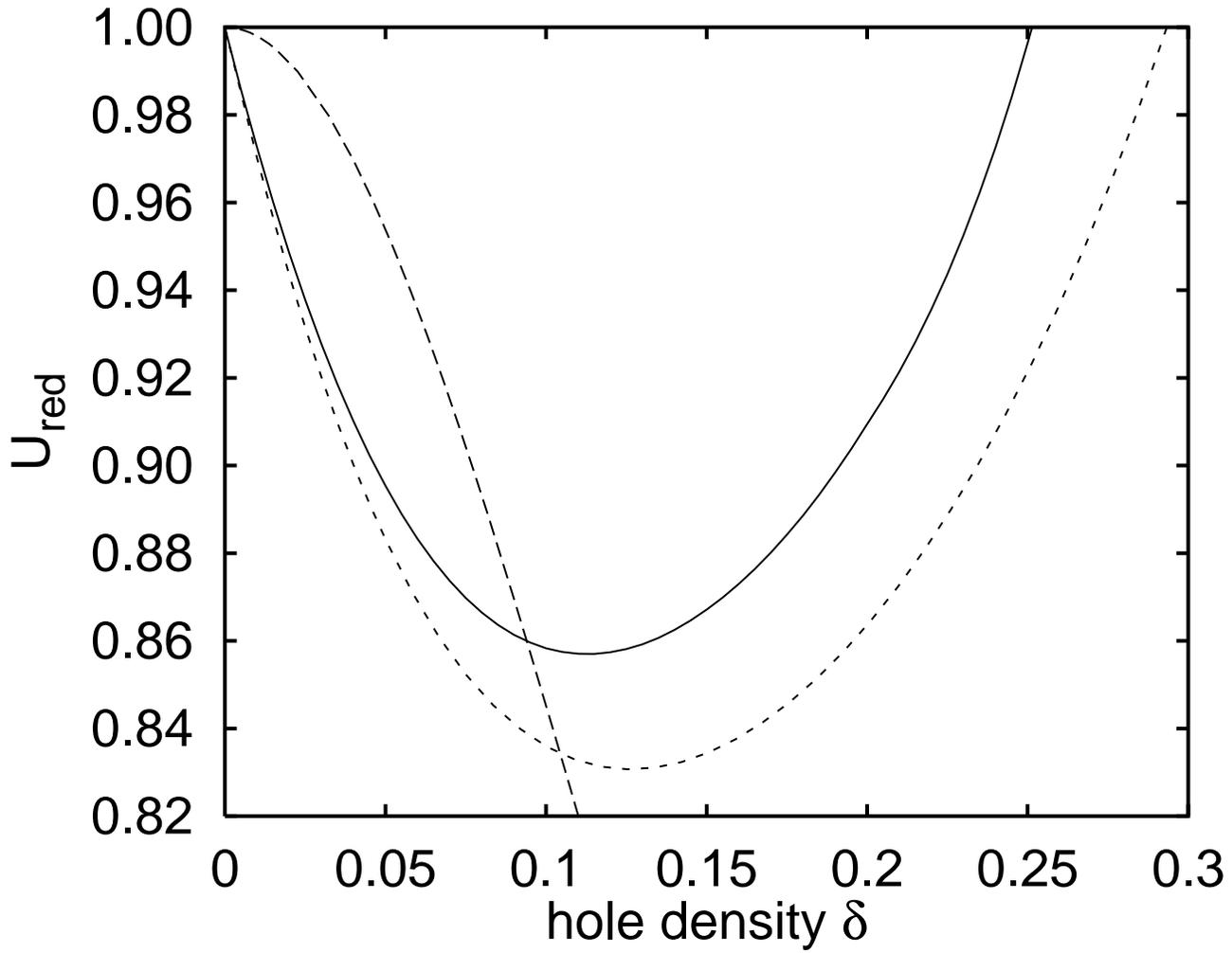

Figure 1: Remaining region of stability for the Nagaoka state, for a wave function with 1100 correlation terms (full line), for the best wave function from [19] with 28 parameters (short dashed line) and for global instability against SDW (dashed line)

11